\documentclass{iaus}
\usepackage{graphicx}

\def\etal{et al.\,}

\def\eg{{\it e.g.,}\,}

\def\la{\hbox{\raise.5ex\hbox{$<$} 
    \kern-1.1em\lower.5ex\hbox{$\sim$}}} 
\def\ga{\hbox{\raise.5ex\hbox{$>$} 
    \kern-1.1em\lower.5ex\hbox{$\sim$}}}

\newcommand{\dgr}{\mbox{$^\circ$}}           
\newcommand{\Msun}{\mbox{$M_\odot$}}         
\newcommand{\Lsun}{\mbox{$L_\odot$}}         
\newcommand{\cm}{\mbox{\ cm}}                
\newcommand{\K}{\mbox{\ K}}                  
\newcommand{\cms}{\mbox{\ cm s${}^{-1}$}}    
\newcommand{\mes}{\mbox{\ m s${}^{-1}$}}    
\newcommand{\gcm}{\mbox{\ g cm${}^{-3}$}}    
 
\title[Hydrodynamic simulations of the core helium flash] 
{Hydrodynamic simulations of the core helium flash}

\author[Miroslav Moc\'ak and Ewald M\"uller]   
{Miroslav Moc\'ak,
 Ewald M\"uller \break Achim Weiss \and Konstantinos Kifonidis}

\affiliation{Max-Planck-Institut f\"ur Astrophysik, 
          Postfach 1312, 85741 Garching, Germany
          \break email: mmocak@mpa-garching.mpg.de}

\pubyear{2008}
\volume{IAU Symposium No.252}  
\pagerange{119--126}
\date{?? and in revised form ??}
\jname{Proceedings Title IAU Symposium}
\editors{A.C. Editor, B.D. Editor \& C.E. Editor, eds.}
\begin{document}

\maketitle

\begin{abstract}
We desribe and discuss hydrodynamic simulations of the core helium
flash using an initial model of a 1.25 $M_{\odot}$ star with a 
metallicity of 0.02 near at its peak. Past research concerned with the dynamics
of the core helium flash is inconclusive. Its results range from a
confirmation of the standard picture, where the star remains in
hydrostatic equilibrium during the flash (\cite[Deupree\,
1996]{deu:96}), to a disruption or a significant mass loss of the star 
(\cite[Edwards 1969]{ed:69}; \cite[Cole \& Deupree\,1980]{cd:80}). 
However, the most recent multidimensional hydrodynamic study 
(\cite[Dearborn \etal 2006]{de:06}) suggests a quiescent behavior of 
the core helium flash and seems to rule out an explosive scenario. 
Here we present partial results of a new comprehensive study of the core
helium flash, which seem to confirm this qualitative behavior and
give a better insight into operation of the convection zone powered by
helium burning during the flash. The hydrodynamic evolution is followed on
a computational grid in spherical coordinates using our new version 
of the multi-dimensional hydrodynamic code HERAKLES, which is based 
on a direct Eulerian implementation of the piecewise parabolic method.
\keywords{Stars: evolution --
                hydrodynamics --
                convection }
\end{abstract}

\firstsection 
\section{Introduction}

First results on the core helium flash were gained from
one-dimensional hydrostatic numerical simulations of a 1.3 
$M_\odot$ star (Z=0.001) (\eg \cite[Schwarzschild \& H\"arm
1961]{sh:62}). During the flash, the star underwent a thermal runaway
due to the ignition of helium under degenerate conditions in its
center. It reached 
a peak at maximum core temperature of $\sim\,3.5\,10^{8}\,K$ and total 
energy generation rate of $\sim\,10^{12} L_{\odot}$. 
The calculations were redone later with better numerical techniques and 
improved treatment of major physical processes (\cite[Sweigert \& Gross
1978]{sg:78}) and although the ignition of helium occured off-center
due to neutrino processes, they did not change the general picture 
mentioned earlier.  It turns out, 
that the typical e-folding times for the energy release from helium burning
become as low as hours at the peak of the flash, and therefore are
comparable to convective turnover times. Thus, the usual assumptions
used in simple descriptions of convection in one-dimensional hydrostatic
calculations (e.g. instantaneous mixing) do not have to be valid any 
longer. Previous attempts to relax these
assumptions by allowing for hydrodynamic flow remained inconclusive
(\cite[Edwards\,1969]{ed:69}; \cite[Deupree\,1996]{deu:96}; 
\cite[Dearborn \etal 2006]{de:06}). Using a modified 
version of the HERAKLES code (\cite[Kifonidis \etal 2003]{ko:03}) 
which is capable of solving the hydrodynamic equations coupled to 
nuclear burning and thermal transport in up to three spatial
dimensions, we want to deepen our understanding of the
convection during the core helium flash at its peak investigating it 
by means of two-dimensional and three-dimensional hydrodynamic 
simulations. 

\section{Initial setup}

\begin{figure*} 
\includegraphics[width=0.49\hsize]{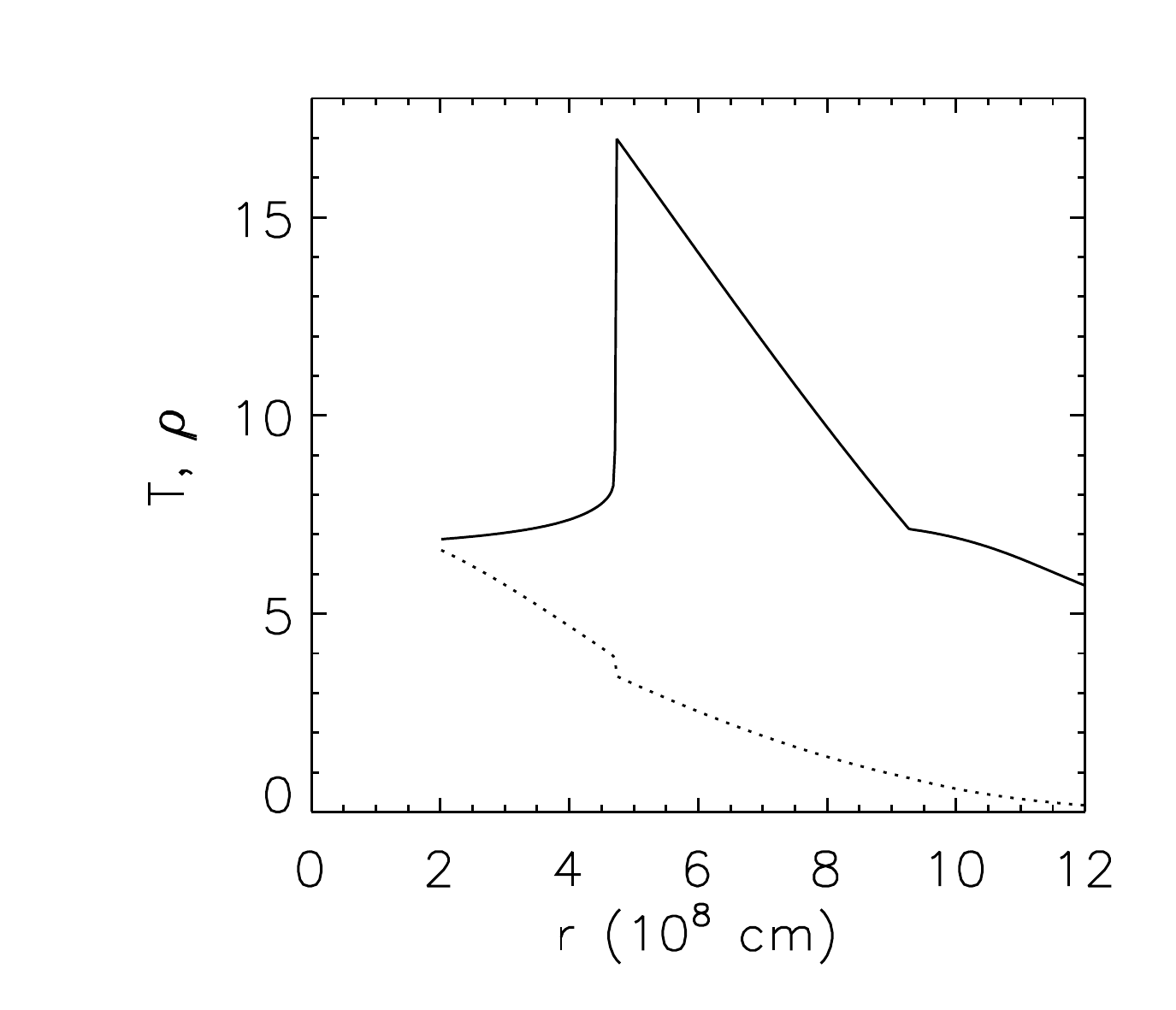} 
\includegraphics[width=0.49\hsize]{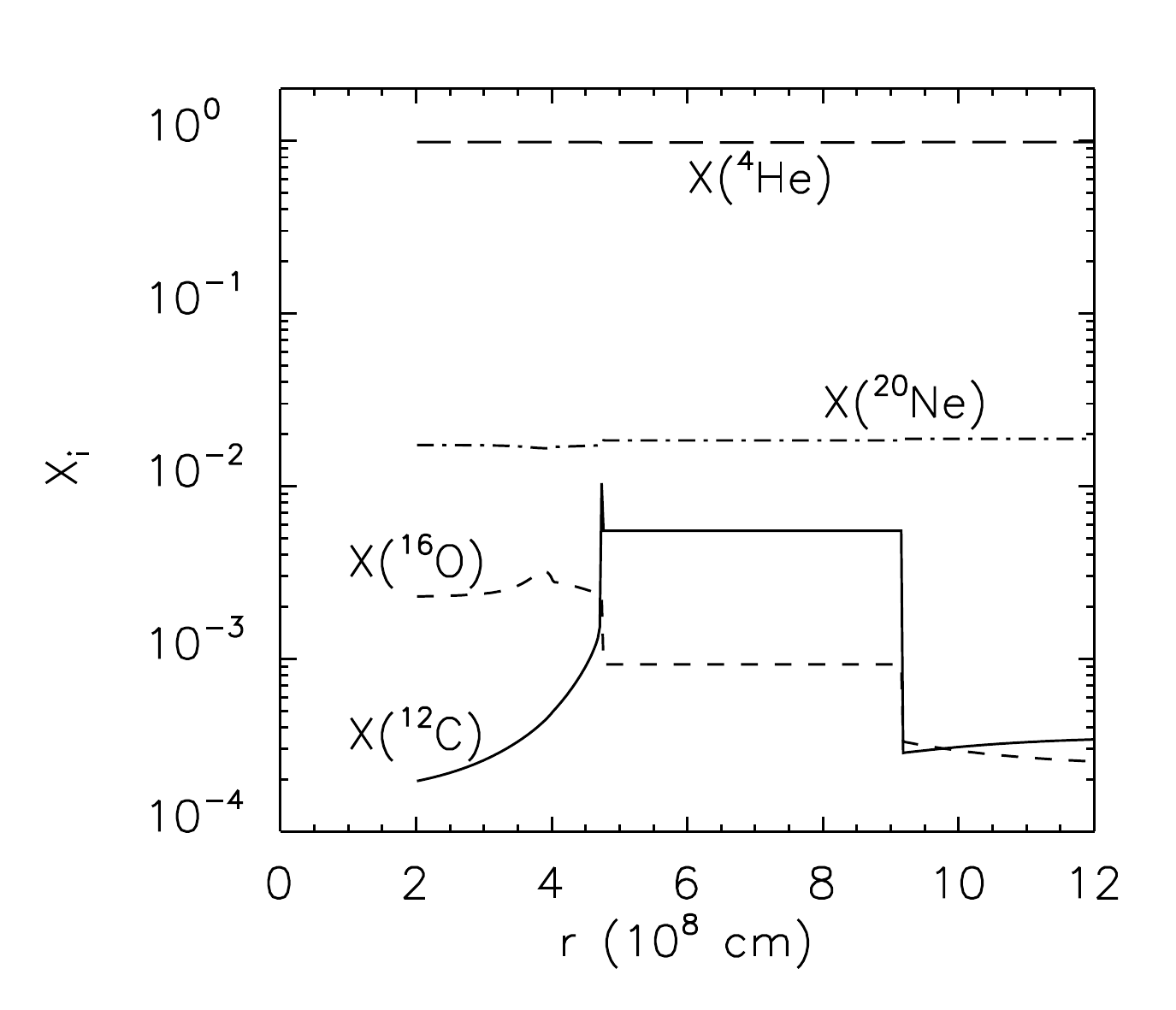} 
\caption{{\it{Left panel:}} Temperature (in $10^{7}\,K$, solid) and density 
(in $10^5\gcm$, dotted) distribution of the initial model M.  
{\it{Right panel:}} Chemical composition of the initial model M, 
showing dominant fraction of helium.}
\label{mmocak.fig1}
\end{figure*}

The initial model was obtained with the stellar evolution code GARSTEC 
(\cite[Weiss \& Schlattl\,2007]{ws:07}). Some of its properties are listed in 
Table\,\ref{mmocak.imodtab}. The temperature, density and composition 
distribution of the model is depicted in Figure\,\ref{mmocak.fig1}. The
model encompasses a white dwarf-like degenerate structure with
an off-center temperature maximum resulting from 
plasma- and photo-neutrino cooling and a central density of about 
$7\,10^5$\gcm. The isothermal region in the center of the helium core 
is followed by almost discontinuous jump in temperature up to $T_{max} \sim
1.7\,10^{8}\,K$ and convection zone driven by the superadiabatic
temperature gradient. The model is composed
mostly of helium $^{4}$He with an abundance X($^{4}$He)$>$ 0.98. 
The remaining composition of the 
stellar model is $^{1}$H, $^{3}$He, $^{12}$C, $^{13}$C, $^{14}$N, 
$^{15}$N and $^{16}$O. For our hydrodynamic simulations we adopt  
the abundances of $^{4}$He, $^{12}$C and $^{16}$O from the initial
model, since the triple-$\alpha$ reaction dominates the nuclear energy 
production rate during the flash. The remaining composition is assumed 
to be adequately represented by a gas with a mean molecular weight 
equal to that of $^{20}$Ne.

\begin{table}\def~{\hphantom{0}}
  \begin{center}
  \caption{Some properties of the initial model: total mass $M$,
           stellar population, metal content $Z$, mass $M_{He}$ and
           radius $R_{He}$ of the helium core ($X(^{4}He) > 0.98$),
           nuclear energy production in the helium core $L_{He}$,
           maximum temperature of the star $T_{max}$, and radius
           $r_{max}$ and density $\rho_{max}$ at the temperature
           maximum.}
  \label{mmocak.imodtab}
\vspace{0.4cm}
\begin{tabular}{l|lllllllll} 
Model & $M$           & Pop.       & $Z$         & $M_{He}$     & $R_{He}$ 
      & $L_{He}$      & $T_{max}$  & $r_{max}$   & $\rho_{max}$ \\ 
      & $[\Msun]$     &            &             & $[\Msun]$    & $[10^9\cm]$ 
      & $[10^9\Lsun]$ & $[10^8\K]$ & $[10^8\cm]$ & $[10^5\gcm]$ \\
\hline 
M  & $1.25$ & I      & $0.02$ & $0.47$ & $1.91$ & $1.03$ 
   & $1.70$ & $4.71$ & $3.44$ \\
\end{tabular} 
 \end{center}
\end{table}

\section{Hydrodynamic simulations}

Table\,\ref{mmocak.modtab} summarizes some characteristic parameters of
our two-dimensional (2D) and three-dimensional (3D) simulations that
are based on model M. They were performed
on an equidistant spherical grid encompassing 95\% of the helium
core's mass except for a central region with a radius of 
r $\,=\,2\,10^8\cm$, which was excised in order to allow for 
larger timesteps.

\begin{table}\def~{\hphantom{0}}
  \begin{center}
  \caption{Some properties of the two and three-dimensional
    simulations: number of grid points in radial ($N_{r}$) and angular 
   ($N_{\theta}, N_{\phi}$) dimension, radial ($\Delta r$ in 
   $10^{8}\,\cm$) and angular ($\Delta \theta, \Delta \phi$)
   resolution, characteristic length scale $l_{c}$ (in $10^{8}\,\cm$) 
   and velocity $v_{c}$ (in $10^{6}\,\cms$) of the flow, respectively, 
   expansion velocity at the position of temperature maximum 
   $v_{exp}$ (in $\,\cms$), entrainment rate $v_{ent}$ of the outer 
   convective boundary (in $\mes$), typical convective turnover time 
   $t_{o}$ and maximum evolution time $t_{max}$ (in s), respectively.}
  \label{mmocak.modtab}
\vspace{0.4cm}

\begin{tabular}{p{1.cm}|p{2.4cm}p{0.7cm}p{0.7cm}p{0.7cm}
  p{0.7cm}p{0.7cm}p{0.7cm}p{0.7cm}p{0.7cm}p{0.7cm}p{0.7cm}} 
\hline
run & $N_{r} \times N_{\theta} \times N_{\phi}$ & 
$\Delta r$ & $\Delta\theta$ & $\Delta\phi$ & $l_{c}$ & 
$v_{c}$ & $v_{exp}$ & $v_{ent}$ & $t_{o}$ & $t_{max}$ \\
\hline 
DV2 & $180\times90$ & 5.55 & 2.\dgr & -  &
4.7 & 1.03 & -\,\,6. & 7. & 910 & 30000  \\
DV4 & $360\times240$ & 2.77 & 0.75\dgr & -  &
4.7 & 1.52  & +92. & 14. & 620 & 60000  \\
TR & $180\times60\times60$ & 5.55 & 1.5\dgr & 1.5\dgr &
4.7 & 0.7 & +6. & 7. & 1340 & 5300\\
\hline
\end{tabular} 

 \end{center}
\end{table}

All our 2D and 3D models undergo initially (t $<$ 1200 s) a 
common evolution where convection sets in after 
roughly 1000~s. During this phase, hot bubbles appear in the 
region where helium burns in a thin shell
(r$\,\sim 5\,10^{8}\,$cm). After $\sim$ 200~s, they cover 
complete height of the convective region and reach a steady state with 
several upstreams (or plumes) of hot gas carrying the released nuclear 
energy away from the burning region, thereby inhibiting a
thermonuclear runaway. 

\begin{figure*}

\centerline{\includegraphics[width=11.6cm]{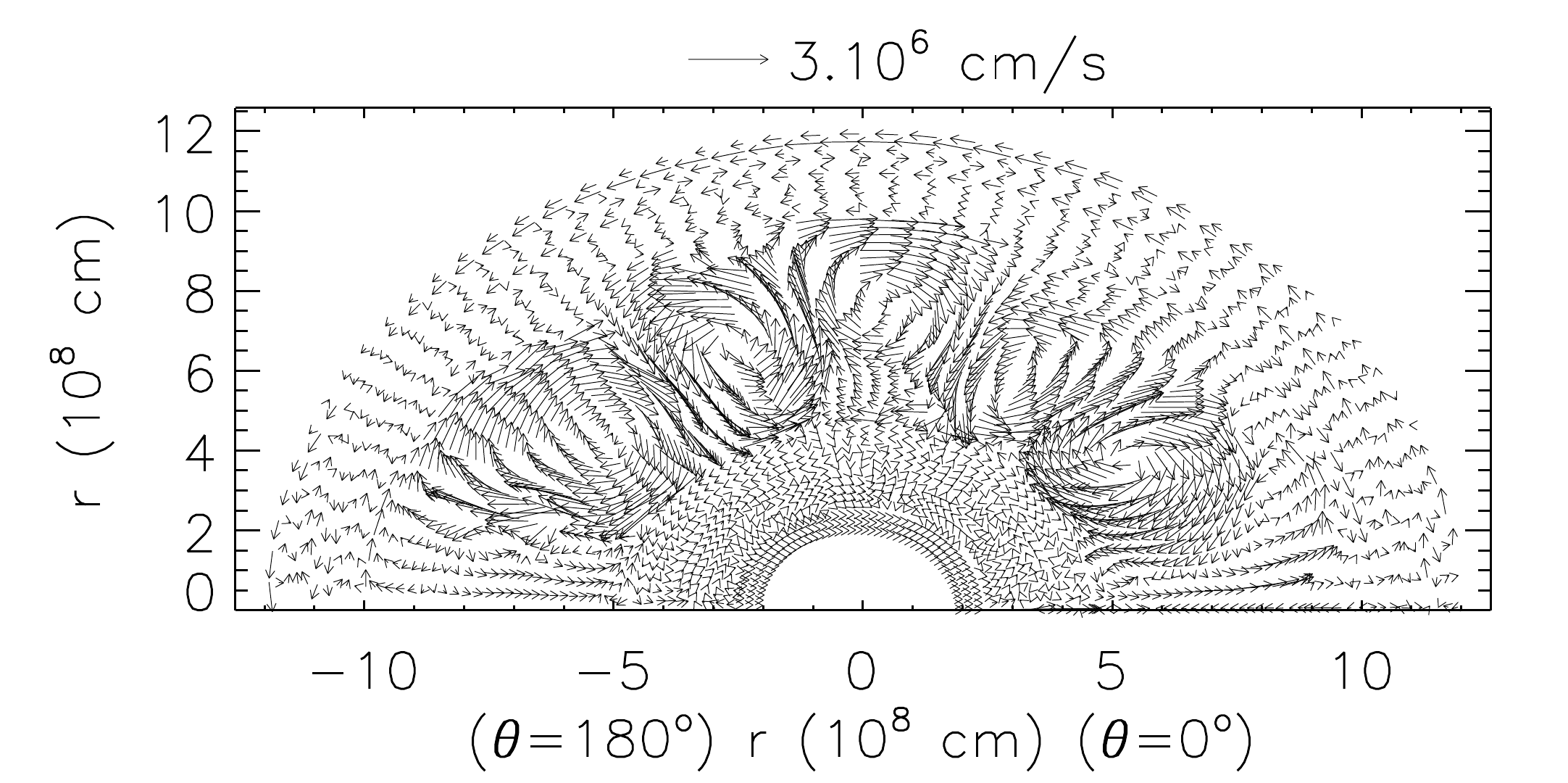}}

\centerline{\includegraphics[width=11.6cm]{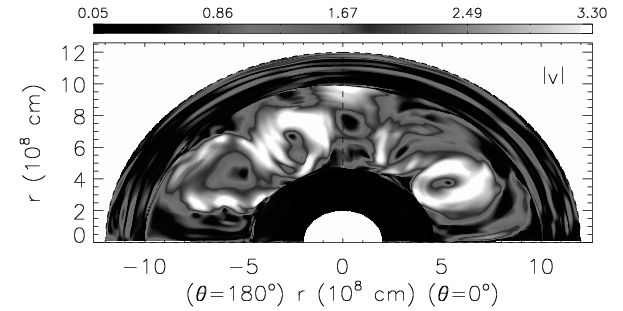}} 

\caption{Snapshots of the evolved convection at $60000\,$s in model
         DV4, showing the velocity field (upper panel), and the
         velocity amplitude $|{\bf{v}}|$ in $10^{6}\,\cms$ (bottom
         panel), respectively.}
\label{mmocak.fig2}
\end{figure*} 

\begin{figure}
\includegraphics[width=6.7cm]{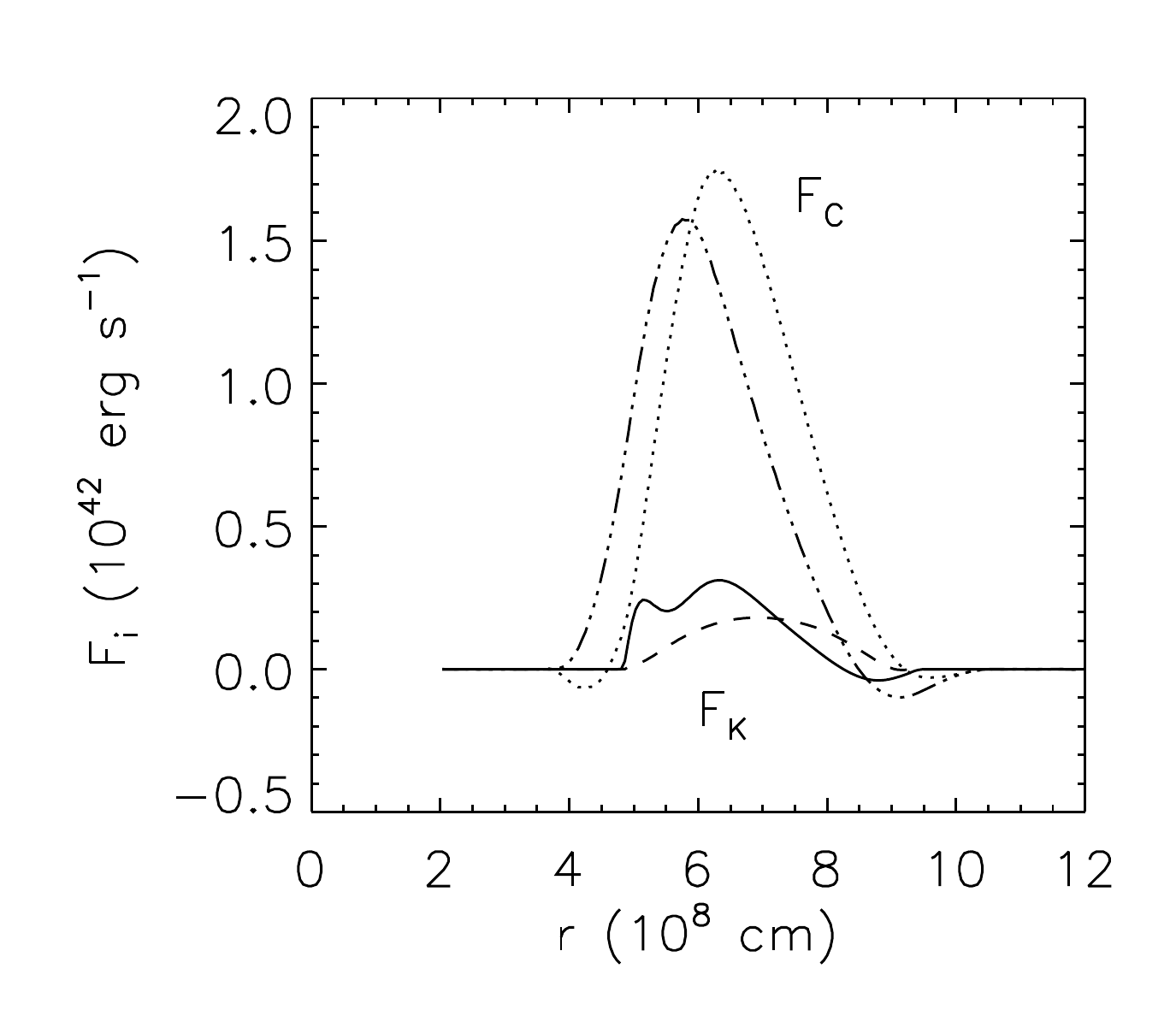}
\includegraphics[width=6.7cm]{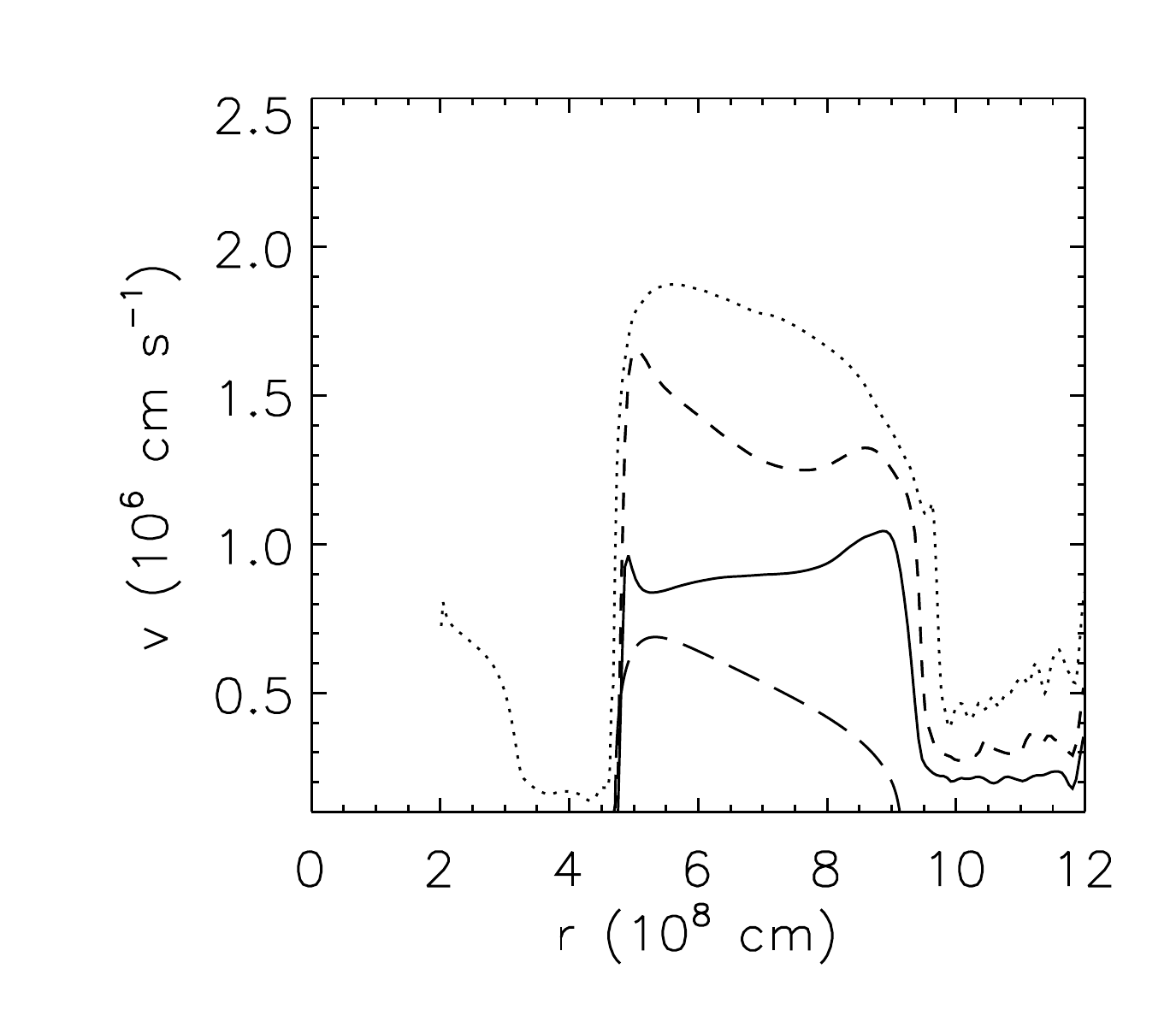}
\includegraphics[width=6.7cm]{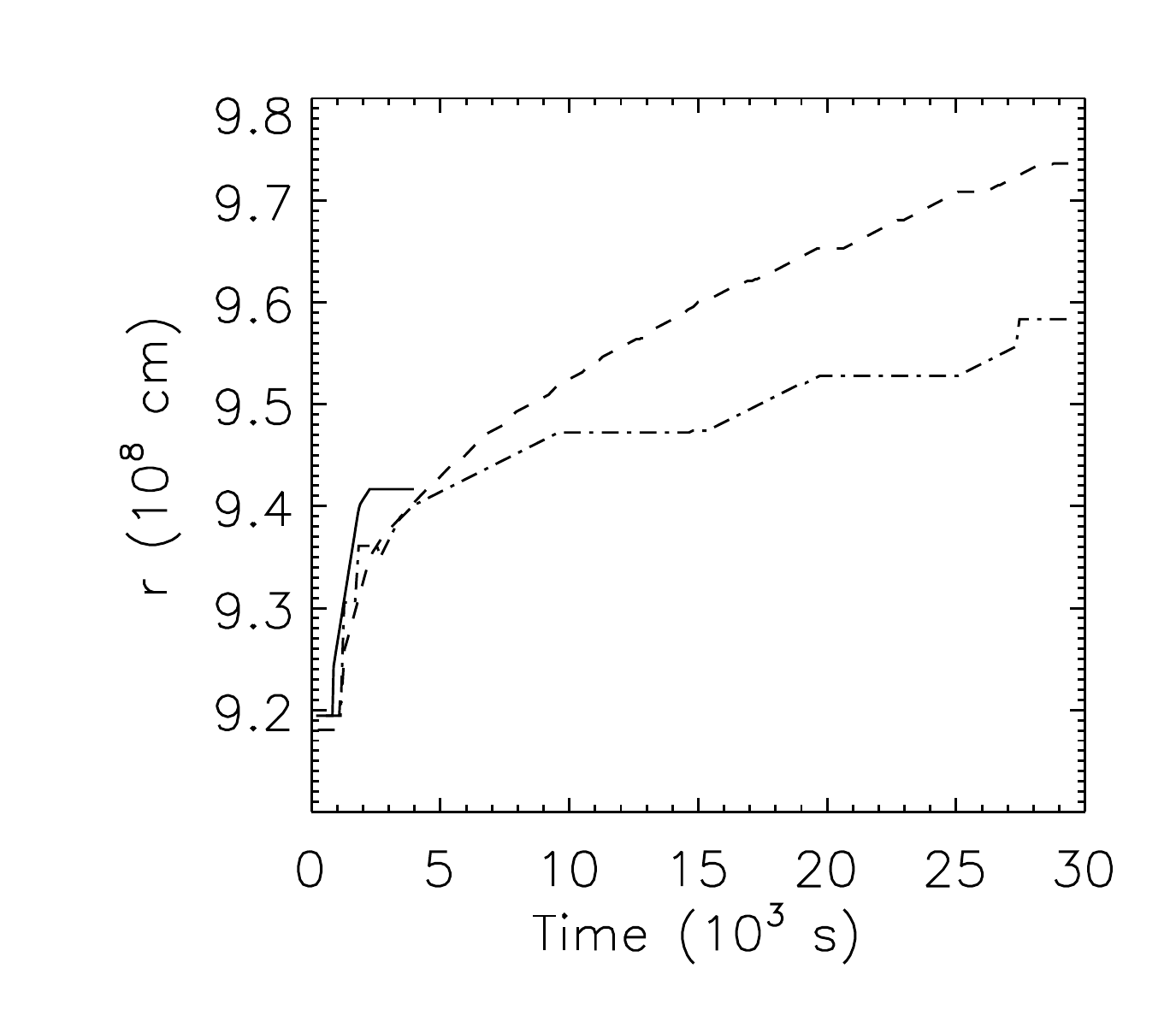}
\includegraphics[width=6.7cm]{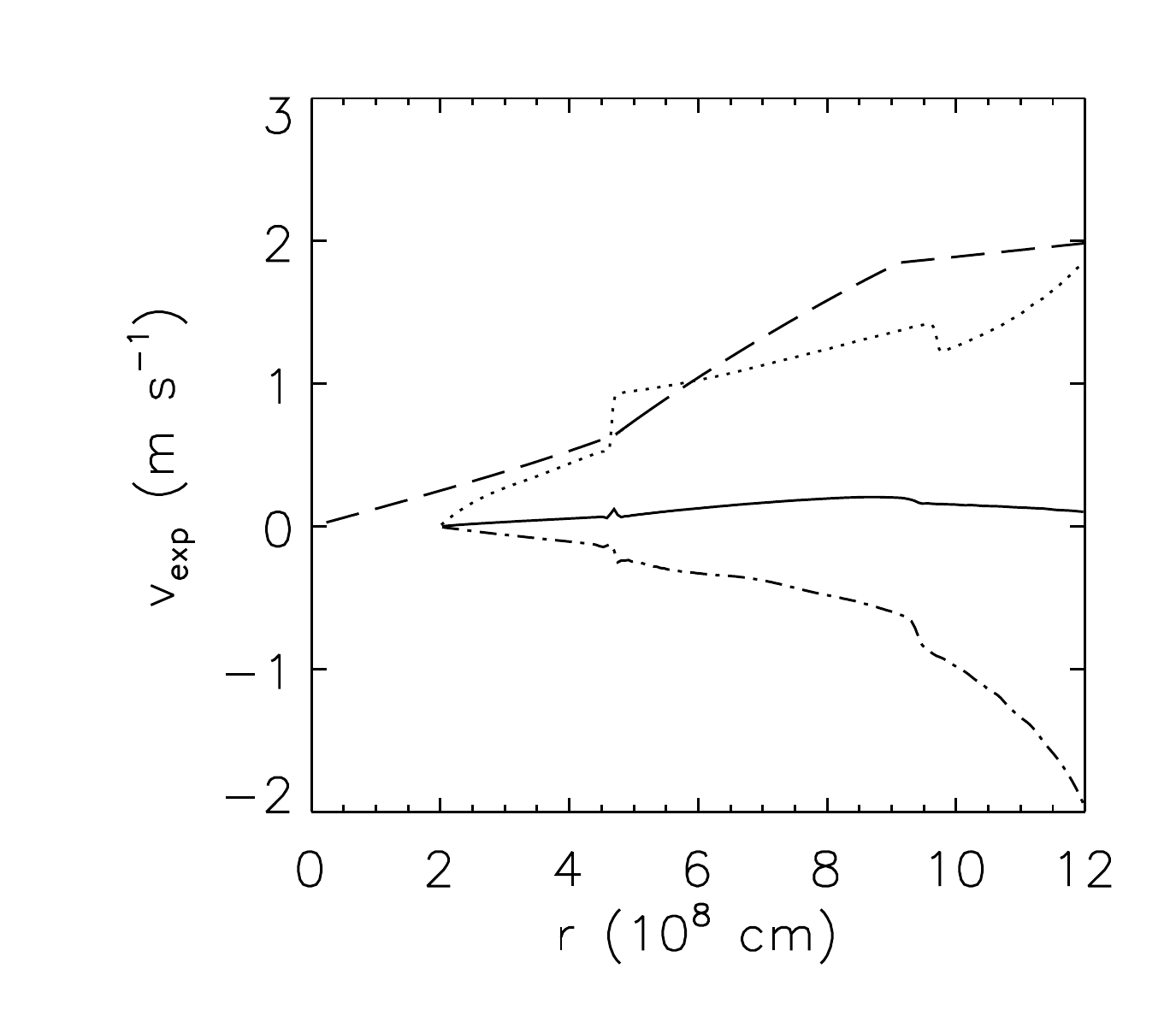}
\caption{{\it{Left upper panel:}} Convective energy flux $F_C$
  in model DV2 (dotted) and model TR (dash-dotted) and kinetic energy 
  flux $F_K$ in model DV2 (solid) and model TR (dashed)  
  (\cite[Hurlburt \etal 1986]{hl:86}). 
  {\it{Right upper panel:}} The r.m.s convection velocity in model
  DV2 (dashed), DV4 (dotted) and TR (solid) overplotted with
  the convective velocities predicted by mixing-length theory
  (long-dashed).
  {\it{Bottom left panel:}} Temporal evolution of the outer convective 
  boundary in model DV2 (dash-dotted), DV4(dashed) and TR
  (solid), respectively.
  {\it{Bottom right panel:}} Expansion velocity 
  v$_{exp}$ in model DV2 (dash-dotted), DV4 (dotted) and TR (solid) 
  together with the expansion velocities of the initial stellar 
  model (long-dashed).}
\label{mmocak.fig3}
\end{figure} 

Fully evolved convection (t $>$ 1500 s) in 3D is significantly 
different than in 2D, since the shape of turbulent streams which 
transport energy is totally distinct. However, the amount of energy 
which needs to be
transported by the convection in order to prevent a thermonuclear 
runaway during the flash is in both cases similar. The resulting typical
convective velocities are therefore much higher in 2D than in 
3D (Fig.\,\ref{mmocak.fig3}). 

\begin{figure}

\includegraphics[width=6.7cm]{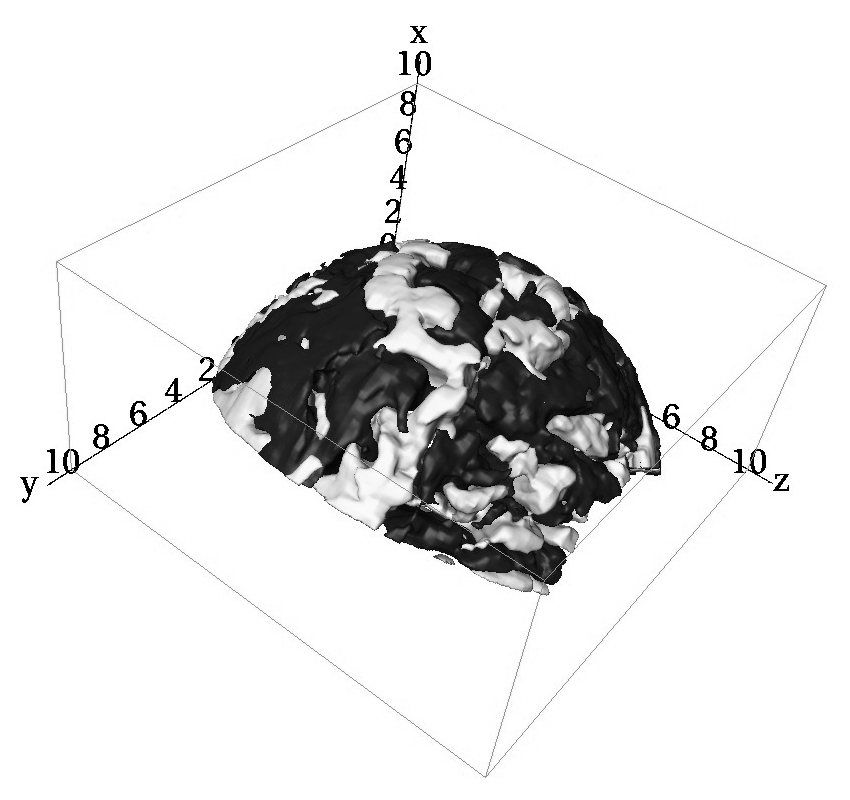}
\includegraphics[width=6.7cm]{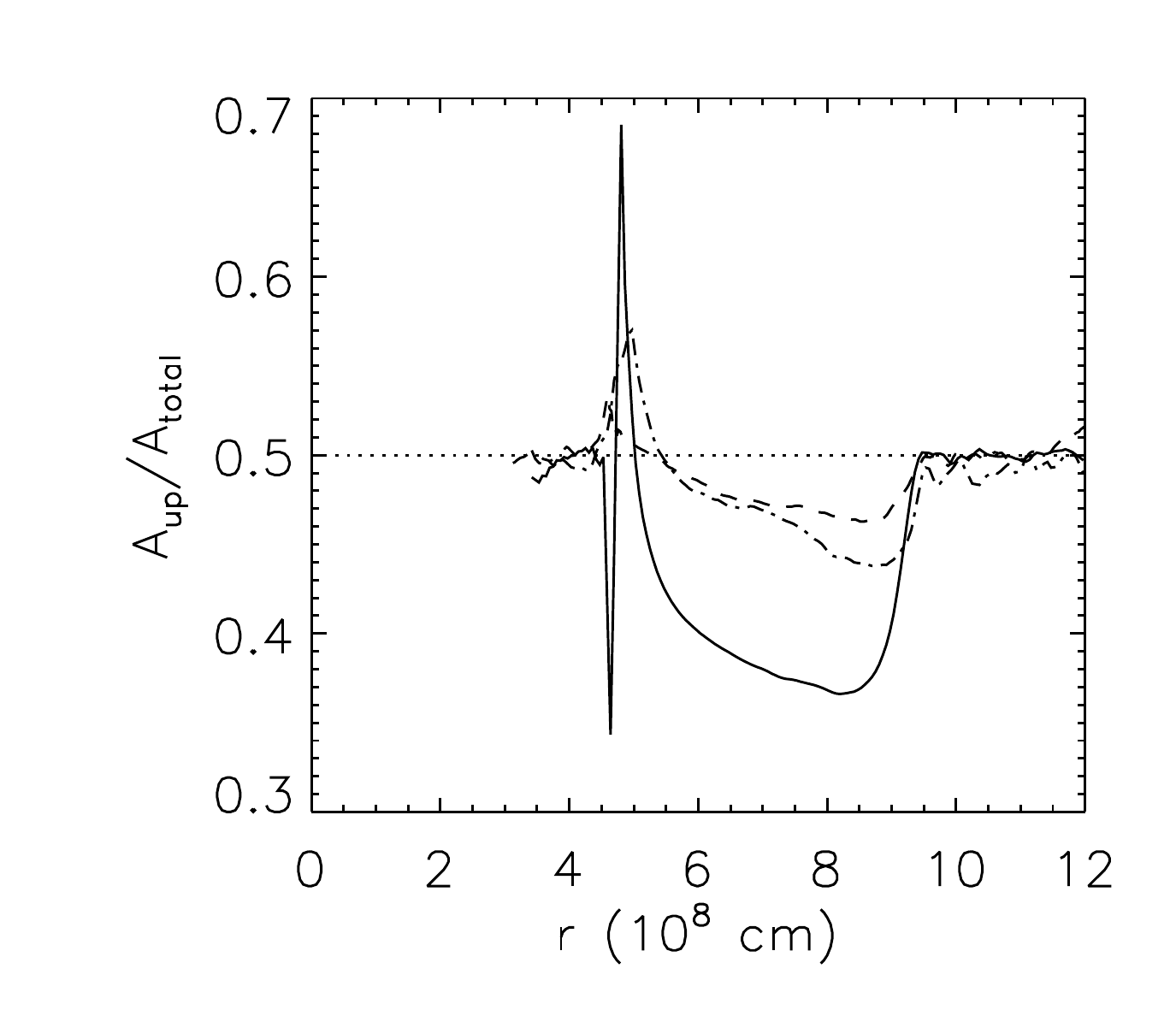}

\caption{{\it{Left panel:}} Isosurface of a radial velocity field of
model TR at t = 4150 s. The bright color marks a positive velocity 
of $+6\,10^{5}\cms$ (the upflow streams) and the dark color a negative
velocity of $-6\,10^{5}\cms$ (downflow streams). Axis tickmarks
are in units $10^{8}\cm$. 
{\it{Right panel:}} Fractional volume occupied by upflow and downflow
streams in model DV2 (dashed-dotted), DV4 (dashed) and 
TR (solid), respectively.}
\label{mmocak.fig4}
\end{figure}

The structural differences between 2D and 3D flows are clearly visible
in the distribution of the kinetic flux across the convection
zone (Fig.\,\ref{mmocak.fig3}). The typical evolved 2D 
flows contain well defined vortices (Fig.\,\ref{mmocak.fig2}) with
their central regions never interacting with the region of the
dominant nuclear burning above the $T_{max}$. This results in a reduced
kinetic flux between $5\,10^{8}\,\cm <$ r $<\,6\,10^{8}\,\cm$, since 
the gas in that region, on average, is located at bottom of convective
vortices, and thus does not experience any strong radial flow. 
On the other hand, 
the distribution of the kinetic flux in 3D is rather smooth, 
and the flow structures tend to be also smaller than in 2D. 
This is apparent when comparing  
Figure\,\ref{mmocak.fig4} with Figure\,\ref{mmocak.fig2}. The 2D 
structures (vortices) have an angular size of around $40\dgr$. The 
structures in the 3D are column shaped, with a smaller angular 
size. The convective and kinetic flux is lower in 3D than in the 2D, but 
the total energy production is about 20 \% higher in 3D (because no 
symmetry restrictions are imposed, and due to the strong dependence of
the triple-$\alpha$ reaction rate on the temperature).
The convective and kinetic flux carry together more than 
90 \% of the energy produced by the burning. The 3D velocities  
qualitatively match the velocities predicted by the mixing-length 
theory better than in 2D, where the velocities are clearly 
overestimated. Figure\,\ref{mmocak.fig3} shows that they also depend 
on resolution, being higher in the simulation with the highest 
resolution.
 
The extent of the convection zone increases with time.
Due to turbulent entrainment (\cite[Meakin \& Arnett 2007]{ma:07}), 
convective boundaries defined by the Schwarzschild criterium are 
pushed towards the center of the star, and towards the 
stellar surface, respectively (Fig.\,\ref{mmocak.fig3}).  This is in
contradiction with the predictions made by (1D) hydrostatic
stellar modeling. For ilustration, the temporal 
evolution of the location of the outer convective boundary is 
depicted in Figure\,\ref{mmocak.fig3}. It is 
defined as the radius, where the mean carbon abundance
X($^{12}$C\,)$\,\sim\,$ 0.002. The rapid initial jump of the
boundary position to r $\,\sim\,9.4\,10^{8}\,\cm$    
at about $\sim$ 1200 s is due to the first touch of the 
convective flow on the boundary. Later entrainment is 
rather steady. The velocity of the outer boundary 
(entrainment rate) in our models are listed in Table\,\ref{mmocak.modtab}.
The entrainment involves a few radial zones only over the longest 
simulation we have performed. Although the 
$^{12}$C abundance distribution stayed discontinuous at boundaries
(no evident effect of numerical diffusion is detected), the entrainment 
rates presented here have to be considered as an order of magnitude 
estimate only. The entrainment at the inner convective boundary occurs 
with a rate much smaller than at the outer convective boundary.
Therefore it is not discussed further here, since longer simulations
are needed for definite statements about its evolution. 

A similar feature which 2D and 3D seem to share is the upflow-downflow
asymmetry. The downflows cover a much bigger volume in the convection
zone than the upflows. The downflows dominate more in
3D. Interestingly, the kinetic flux is always positive in both cases, 
although the downflows fill almost the whole convection zone. 
This implies that the downflows are much slower then the upflows. 

The expansion velocities v$_{exp} = \dot{M_{r}}/4 \pi r^{2} \rho$
are in good agreement with those of initial
stellar model only in the 2D model with highest resolution DV4 
(Fig.\,\ref{mmocak.fig3}). The expansion in the low 
resolution models does not match at all. Due to the different dynamic 
properties of the flow in the less resolved models, the spherical mass 
flow is weaker. 

\begin{figure}
\includegraphics[width=4.4cm]{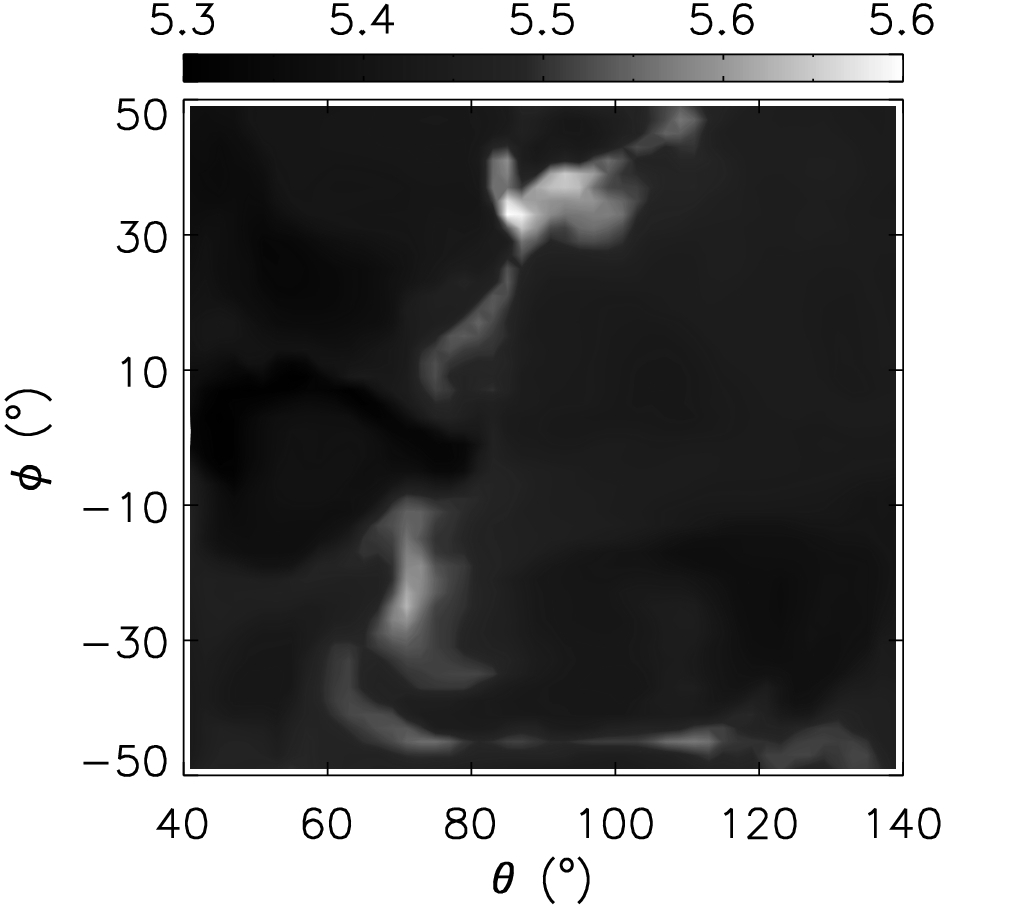}
\includegraphics[width=4.4cm]{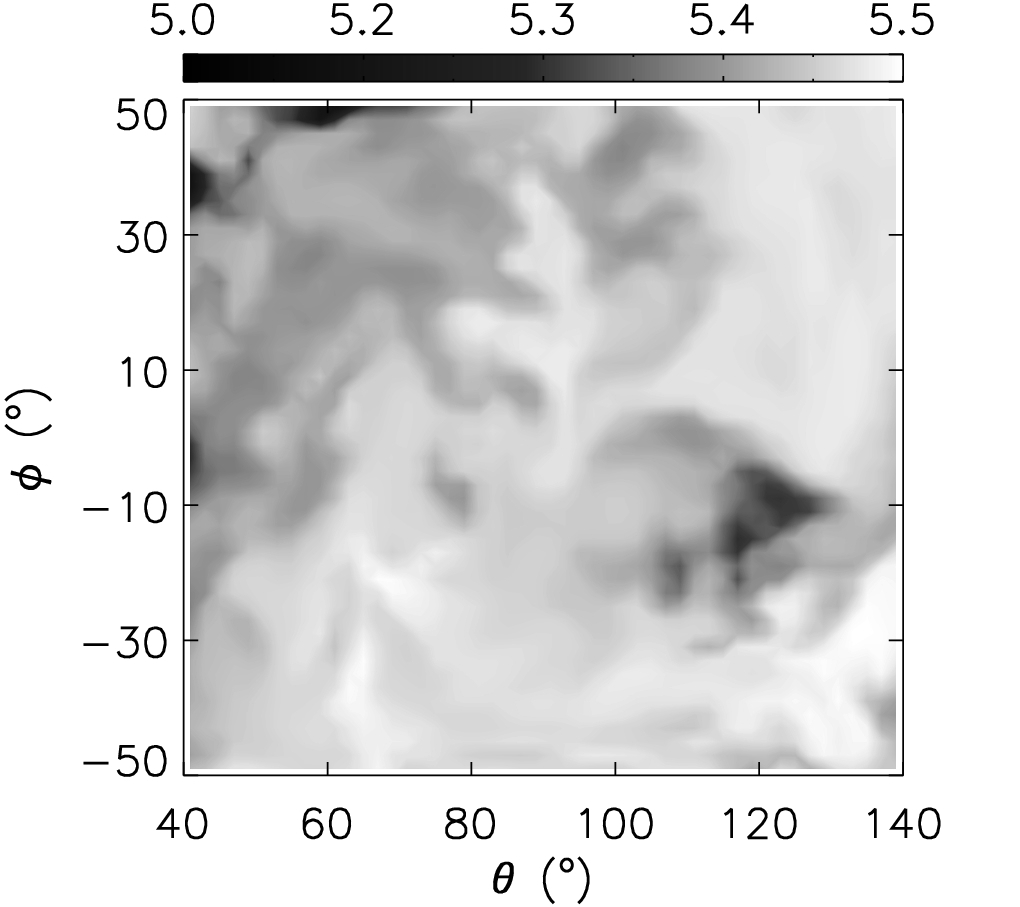}
\includegraphics[width=4.4cm]{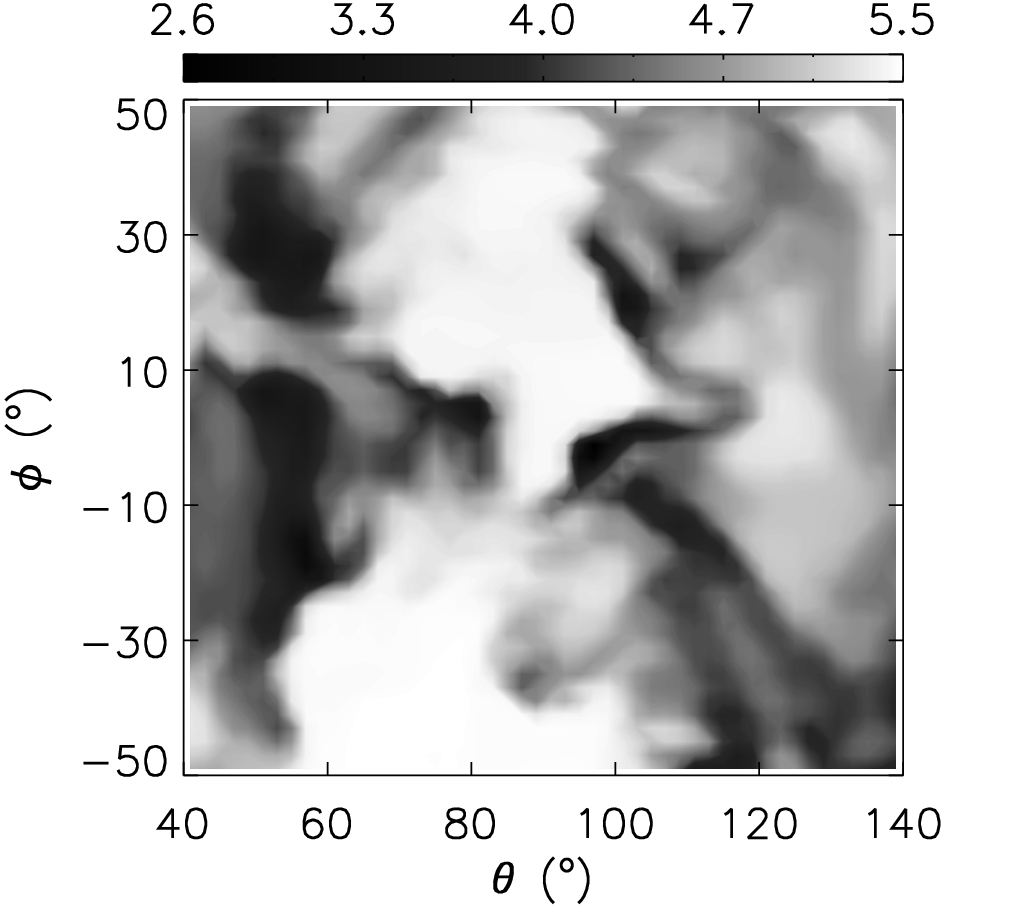}
\caption{Maps of $^{12}$C abundance (in units of $10^{-3}$) in a 
horizontal plane in model TR at t = 5231 s, at different radii: 
r$_{1}$ = 4.8$\,$10$^{8}$cm (left), r$_{2}$ = 6.5$\,$10$^{8}$cm (middle), 
r$_{3}$ = 9.3$\,$10$^{8}$cm (right).}
\label{mmocak.fig5}
\end{figure}

Mixing at the convective boundaries and across the convection
zone in 2D and 3D is quite different as well. In 2D, due to the 
symmetry restriction, every turbulent feature is in fact an 
annulus. Hence, turbulence and mixing can
be properly modelled only by means of 3D simulations. The most
apparent turbulent structures during the flash, in 3D, look at the 
bottom of the convection zone like thin hot fibers enriched by carbon 
and oxygen (ashes from the helium burning). The flow then gets more 
uniform inside the convective region, but looks more turbulent again 
at the outer convective boundary (Fig.\,\ref{mmocak.fig5}).   

\section{Conclusions}

We find that the core helium flash neither rips the star apart,
nor significantly alters its structure. The evolved convection in 
3D looks different from that in 2D.
Typical convective velocities are higher in 2D than in 3D where 
they also tend to fit the predictions made by mixing length theory 
better. Hydrodynamic simulations show the presence of turbulent 
entrainment, which results in a growth of the convection zone on 
dynamic time scales.

\begin{acknowledgements}
The calculations were performed at the Leibniz-Rechenzentrum of the
Bavarian Academy of Sciences and Humanities on the SGI Altix 4700
system. A considerable grant of computer time is thankfully acknowledge.

\end{acknowledgements}

\end{document}